\documentstyle[12pt,aasms4]{article}

\begin{document}

\title{A major star formation region in the receding tip of the stellar
Galactic bar. \\ II. Supplementary information and evidence that the
bar is not the same structure as the triaxial bulge previouly reported.}

\author{M. L\'opez-Corredoira}
\affil{Instituto de Astrof\'{\i}sica de Canarias, E-38200 La Laguna,
Tenerife, Spain}
\author{F. Garz\'on}
\affil{Instituto de Astrof\'{\i}sica de Canarias, E-38200 La Laguna,
Tenerife, Spain \\
Departamento de Astrof\'\i sica. Universidad de La Laguna, E-38204 La Laguna,
Tenerife, Spain \\
}
\author{J. E. Beckman} 
\affil{Instituto de Astrof\'{\i}sica de Canarias, E-38200 La Laguna,
Tenerife, Spain \\
Consejo Superior de Investigaciones Cient\'\i ficas. Spain}
\author{T. J. Mahoney}
\affil{Instituto de Astrof\'{\i}sica de Canarias, E-38200 La Laguna,
Tenerife, Spain}
\author{P. L. Hammersley}
\affil{Instituto de Astrof\'{\i}sica de Canarias, E-38200 La Laguna,
Tenerife, Spain}
\author{X. Calbet}
\affil{Instituto Nacional de Meteorolog\'\i a, Santa Cruz, Tenerife, Spain}

\begin{abstract}

This paper is the second part of Garz\'on {\it et al.} (1997) in which
we presented an outline of the analysis of 60 spectra from a
follow-up program to the Two Micron Galactic Survey 
(TMGS, Garz\'on {\it et al.} 1993) 
project in the $\ell=27$\arcdeg, $b=0\arcdeg \ $ 
area. In this second part, we present a more detailed explanation of
the analysis as well a library of the spectra for more complete
information for each of the 60 stars,
and further discussions on the implications for the structure of the Galaxy.

This region contains a prominent excess in the
flux distribution and star counts previously observed in several
spectral ranges, notably in the TMGS. 
More than 50\% of the spectra of the stars detected with $m_K<5.0$ mag, 
within a very high confidence level, correspond to stars of luminosity
class I (Garz\'on {\it et al.} 1997), and a significant proportion of the remainder are very late
giants which must also be rapidly evolving. 
We make the case, using all the available evidence, that we are 
observing a region at 
the nearer end of the Galactic bar, 
where the Scutum spiral arm breaks away, and that
this is powerful evidence for the presence of the bar.
Regions of this type can form due to the concentrations of
shocked gas where a galactic bar meets a spiral arm, as is observed at
the ends of the bars of many face-on external galaxies. 
Alternative explanations do not give nearly such a satisfactory account of the 
observations.

Equivalent spectroscopical analysis should also be performed 
at $\ell=-22\arcdeg $, the candidate position for the other tip 
of the bar. The space localization of one and, {\it a fortiori}, of both ends
of the bar allows us to infer its orientation. 
If the second region is also confirmed to be  a powerful star formation region
this would imply a position angle for the bar of 
$\sim 75$\arcdeg \ with respect to the Sun--Galactic centre line.
This geometry is indeed compatible with the range of
distances that we have obtained for the star-forming region at $\ell=27\arcdeg $
from spectroscopic parallaxes.
However, the angle is different from that given by other
authors for the bar and this, we think, is because they refer
to the triaxial bulge and not to the bar as detected here.

\end{abstract}


\section {Introduction}

This paper is the second part of Garz\'on {\it et al.} (1997) 
(hereafter G97) in which
we presented an outline of the analysis of 60 spectra from a
follow-up program to the TMGS project in the $\ell=27$\arcdeg, $b=0\arcdeg \ $ 
area. This analysis was based on the luminosity 
classification of stars using the Ca{\sc ii} 
triplet (Jones {\it et al.} 1984; D\'\i az {\it et al.} 1989), and a 
discussion of this region was presented which strengthens the 
identification of this zone as a star formation region at the 
receding tip of the stellar bar.

In this second paper, we present the spectra, offer 
a more detailed explanation of
the analysis, and supply a more complete
information for each of the 60 stars, extended bibliography,
and further discussions of the physical morphology.

\section{Observations and data reduction}

As explained in G97, 70 spectra of
stars situated in the region around $\ell=27\arcdeg $, $b=0\arcdeg $
selected from the TMGS 
with $m_K<5.0$ mag were taken with the Isaac Newton Telescope in La
Palma (Canary Islands, Spain). 
From these, we had to reject ten
which could not be identified as visible counterparts of TMGS sources.
After cross-correlating the original TMGS source positions
with their {\it Guide Star Catalogue} counterparts and then calculating the
errors in position, we inferred that
TMGS catalog has a positional error box of around 
4 arcsec in right ascension by 7 arcsec in declination. 
The larger error in declination compared with that
in right ascension is due to the form of detector array
used in the survey (Garz\'on {\it et al.} 1993).
This is added to the pointing error of the INT telescope itself 
and implies that there is a non-negligible risk
of taking the spectrum of the wrong star
in such a very crowded area of the sky.
By taking spectra of all the bright stars
within the error box, we could decide {\it post hoc} which 
candidate was the TMGS object with $m_K<5.0$ mag from the spectral 
type and features and $I$-magnitude estimate. In the ten cases mentioned above, 
we could not find any good candidate so these were rejected.

In Table \ref{Tab:espectros} and  Fig. \ref{Fig:coords}, coordinates 
of all the stars in our sample  from the TMGS database are given. 
Investigators who may wish
to use them should be aware that, due to the error of the
TMGS coordinates described above, and the extremely crowded nature
of the field there is often more
than one visible candidate for the infrared source. As we did not use 
more precise coordinates for a given source, we have not needed  to
sharpen the precision of these positions.

\begin{figure}[htb]
\label{Fig:coords}
\end{figure}

The extraction of the spectra, the sky subtraction and the wavelength
calibration were performed using the {\sc iraf} tasks  {\tt apextract}
and {\tt onedspec}  after performing the bias subtraction and the
flat fielding with {\tt ccdred}. The flux was
normalized rather than calibrated.

\section{Luminosity classification}
\label{.lumclclas}

Spectra in the zone of the IR Ca {\sc ii} triplet at 8498.02 \AA, 8542.09 \AA, 
and 8662.14 \AA, present in stars of spectral types later than F5, were examined.
The triplet was chosen as an optimum differentiator between stars of
different luminosity classes. In
earlier types, the Paschen hydrogen lines 
P13 (8665 \AA ), P15 (8545 \AA ) and P16 (8502 \AA)
severely contaminate the
spectral region of interest. For spectral types later than M3 or M4, TiO absorption bands
mask the Ca {\sc ii} triplet, making its features more difficult
to measure, as we will see later.

Measurement of the equivalent width of the Ca {\sc ii} triplet lines was 
carried out according to the prescriptions 
of D\'\i az {\it et al.} (1989) when TiO-band
contamination is not present. A continuum $I_\lambda^{\rm cont}$ is defined
by a linear fit to the median value of two chosen side-bands centered at
8455 \AA \ and 8850 \AA, respectively, with a width of 15 \AA\   and the
equivalent width was obtained by integrating:

\begin{equation}
W_\lambda=\int_{\lambda-15{\rm \AA}}^{\lambda+15{\rm \AA}}\left(
1-\frac{I_{\lambda '}}{I_{\lambda '}^{\rm cont}}\right) d\lambda '
,\label{eqw}\end{equation}
where $I_{\lambda }$ is the measured intensity. 

We also followed D\'\i az {\it et al.} (1989) in defining

\begin{equation}
{\rm EW}=W_{8542 \ {\rm \AA}}+W_{8667 \ {\rm \AA} }
\label{EW}\end{equation}
as the indicator of luminosity class.
The shortest-wavelength line of the
triplet was not used because it is too feeble compared with the neighboring
features and yields larger errors
in the precise measurement of the total Ca{\sc ii} triplet.
Selected reference stars were used to check our EWs and these were the same, 
within the error margin, as those given by D\'\i az {\it et al.} (1989)
for the same objects.    

The Ca{\sc ii} triplet has been used for many years 
as an indicator of luminosity class ({\it e.g.}, Merrill 1934).
More recently Jones {\it et al.} (1984), and
D\'{\i}az {\it et al.} (1989) have calibrated empirically 
the relationship between the
equivalent width  of the Ca {\sc ii} triplet and the luminosity
class for spectral types ranging from F5 to M3.
These authors found some dependence of EW on both metallicity and
temperature, but these effects were much weaker than the dependence on surface
gravity, especially for supergiants.  We have followed their results and adopted their criteria in
assigning luminosity classes from the measured EWs. 
From D\'\i az {\it et al.} (1989), 
if metallicity dependence is assumed negligible, then
\begin{equation}
\log g=7.75-0.65{\rm EW}
\label{logg}
.\end{equation}
On this basis, a source is taken to be a supergiant 
($\log g$ less than approximately 1.75) if EW $>9$ \AA , independently of
metallicity and temperature.
Idiart {\it et al.} (1997) have demonstrated the presence of
correlations of EW with metallicity and temperature, but only
for old stars ($>$ 1 Gyr), and therefore not for supergiants. 
There are also other possible luminosity-class indicators. 
In the wavelength range
we used, the existence of the CN band (7916--7941 \AA ) is a
characteristic of supergiants (MacConnell {\it et al.} 1992), but the Ca {\sc ii} triplet is less affected 
by noise.

Thirty-eight stars in the sample were not contaminated by the
TiO band, and we used the method described above to obtain their EWs in 
this subsample (see Table \ref{Tab:espectros}). 

The remaining 22 stars are of later spectral types, and the presence of 
TiO bands partially
masks the triplet lines (see Fig. \ref{Fig:spectraplot} {\it c}), {\it d})). 
For these objects we have developed an
empirical method which permits the measurement of EW where the
Ca {\sc ii} lines are not completely masked by the TiO band. This
method uses the depth of the lines instead of the EW.

The depth of the lines is calculated according to

\begin{equation}
{\rm depth}_\lambda =1-\frac{{\rm min}(I_{\lambda '}; 
\lambda - 15 \ {\rm \AA}<\lambda '<\lambda + 15{\rm \AA} )
}{I_{\lambda }^{\rm cont2}}
,\label{depth}\end{equation}
where $I_{\lambda }^{\rm cont2}$ is the value of a new continuum
interpolated between two points that are not affected by the
TiO absorption: 8432 \AA \ and 8844 \AA \ \footnote{In fact, a bandwidth of 15
\AA \ between 8417 \AA \ and 8432 \AA \ is selected for which we calculate 
the median value, and the same is done for
the second value of the continuum between 8844 \AA \ and 8859 \AA .}.

One potential problem is that the convolution of the profiles of
the spectral lines with the instrumental profile would affect the depth of
the line but this effect is in practice negligible compared with the other
uncertantities
in the method. The convolution with the instrumental profile broadens the line
and decreases the depth when the number of pixels defining the line is small. 
In the lines without TiO contamination, the pixels defining a line cover a 
sufficient wavelength range for
the effect of the convolution to be negligible. In the lines with TiO
contamination, the depth
is the sum of the TiO band depth and the jump between the TiO band and the
peak of the line. If the unblended portion of the line
covers a restricted wavelength range, this jump will be affected by the convolution but,
as it is small compared with the TiO depth, the effect of
the convolution is again small. The effect of convolution would indeed be serious if
we were to use Gaussian fits on masked portions of the lines instead of the depth
as the measure of equivalent width.

To get the EW as a function of the depth of the deepest line, 8542 \AA , it is
necessary to use uncontaminated lines.
Expression (\ref{depth}) is used for the first 38 stars, those with
unmasked Ca{\sc ii} lines, and for these 
a relationship is then calibrated
between depth$_{8542}$ and the EW calculated using eq. (\ref{EW}).
The relationship obtained, employing a linear fit, was:

\begin{equation}
{\rm EW}=-0.1+17.8\times {\rm depth}_{8542}
,\label{EWED}\end{equation}
which is plotted, together with the data, in
Figure \ref{Fig:figEWDW} \footnote{The fit to a
second-degree polynomial  was EW $=-7.1+45.0\times {\rm depth}_{8542}-25.8
\times {\rm depth}_{8542}^2$ which is very close to the linear fit and
yields the same values, within the error limits.  We emphasize that in
G97 we used a second-degree fit, but with a small error, which
gave rise to EW values too low by 0.2 or 0.3 \AA . This led to
our classifying two stars as giants instead of supergiants. This improved
result only strengthens our previous conclusions about the proportion
of supergiants in the sample.}. 
The use of this expression to evaluate the
luminosity class using the criterion of D\'\i az {\it et al.} (1989) 
is a fair approximation. The fit in equation (\ref{EWED}) is a good
approximation but in the latest spectral types other effects may intervene.
The equivalent width of the Ca~{\sc ii} triplet is not strongly dependent 
on spectral type, although this dependence is a 
little more noticeable for M-type stars
since the ionization equlibrium in these objects shifts somewhat from Ca~{\sc ii} 
to Ca~{\sc i} (Cohen 1978).
One consequence is that this line-depth technique may slightly underestimate
the number of supergiants. That is, the number of supergiants among stars
later than M3 may well be in fact larger than that inferred
using the  EW $>9$ criterion applied to equation (\ref{EWED}).
Since our purpose is to calculate the fraction of supergiants 
to determine whether the region studied is a star-formation
region, this underestimate could change our
conclusion only if the answer were negative, but this is not the
case, as will be shown below.

\begin{figure}[htb]
\label{Fig:figEWDW} 
\end{figure}

Even employing the line-depth technique, we had to reject two stars of the sample 
for which the TiO band
gave an unacceptable blend, so we finished with 58 stars---38 earlier 
than M3, and 20 later than M3---with acceptable EWs. 
The two stars rejected show spectra with no clearly unblended Ca{\sc ii}
features; the Ca~{\sc ii}+TiO blends show a 
maximum depth$_{8542\ \AA}$ from which we may estimate
a maximum EW. In both cases, this maximum value of EW is $\approx $10.2 \AA ,
i.e. EW $<10.2$, so it could not be decided whether they are giants
(EW $<9.0$) or supergiants (EW $>9.0$). In Table \ref{Tab:espectros},
we do show the data for these two stars with EW $<10.2$ (stars 29 and 37) but these are
not included in either the statistics of the EW or in the distance
calculation of paper G97.

The final results of the luminosity classification were shown in Fig. 1
of G97, in the form of a histogram of EW frequencies. 
Those EW values used for the figure are shown in Table \ref{Tab:espectros} 
for each star.

The ratio of SGs---those with EW $> 9$ \AA---to the total
number of observed objects is most striking. 
The number of SGs is in fact 38 out of 58 (66\%).
In a binomial distribution the root mean square of the distribution is
$\sigma=\sqrt{n\times p \times (1-p)}$,
where $n$ is the total number of stars. Here, $n=58$ stars,
$p=0.66$, so $\sigma =3.6$ stars. The proportion of supergiants is then
$66\pm 6$\% (1 $\sigma $). An error should be added to allow for 
the possible mistaken classification of 
giants and supergiants, and also to take into account 
the pointing error of the telescope in a crowded field which gives a slight
chance of taking the spectrum of a wrong star.
However, if the ``wrong'' star turned out to be a late star with
sufficient luminosity to give a spectrum of bright star, this would still
be indication of late type giants and supergiants in the zone.
These sources of error are small
except in the  M-type classification, in which the number of supergiants
should be even greater that given by our criterion, as discussed before. 
These considerations lead to the conclusion that
the fraction of supergiants is well over 50\% with a very high confidence level.

\section{Spectral classification}

The spectra of the 60 stars belonging to
both subsamples are presented in Fig. \ref{Fig:spectraplot}, 
after removing the intrinsic slopes and those due to reddening, in
order to show the qualitative differences in the spectra between different
kinds of stars. The spectral type classification 
is carried out for every star by qualitatively comparing the features of
our spectra with those of standard stars. 
The spectral classification is shown in Table \ref{Tab:espectros}. 

The coolest stars---M3 and later on---were compared with those in Bessel (1991), 
Schulte-Ladbeck (1988), and Barbieri {\it et al.} (1981), by inspecting 
the depth of the TiO band absorption. 
In Fig \ref{Fig:spectraplot} {\it c}) and {\it d}), it
can be seen that there is a growing depression at 8432 \AA \ and
8844 \AA \ from M3 to M9. 
Since the features of these sources 
permit us to differentiate spectral types with 
an interval of 0.1 or 0.2 spectral types, the classification is quite
accurate for these stars.

The earlier-type stars were compared with
those in Torres-Dodgen \& Weaver (1993), and Schulte-Ladbeck (1988), mainly
by inspecting lines such as Mg~{\sc i} 8807 \AA , Fe~{\sc i} 8612 \AA , 
Fe~{\sc i}--Ti~{\sc i}
8468 \AA , O~{\sc i}--Fe~{\sc i} 8446 \AA \ (some of these features are also dependent on luminosity
class or metallicity) and Paschen hydrogen lines 
for those earlier than G2. In this case
the errors in the classification could be as
large as half a spectral type, except perhaps for those containing
Paschen lines. 

A histogram of the frequency of spectral types in our sample is shown
in Figs. 2 and 3 of G97. 
As expected, most of the stars are very red and with intrinsic luminosity
(see also Calbet {\it et al.} 1996{\it a}) 
since TMGS is more sensitive to redder stars.
The predominant type among the supergiants is K.
Supergiants of very late types 
are known to exist in very small numbers;
we have found a significant number of new examples here.
This result is subject to the bias that our method of predicting
the EWs gives where the TiO band affects the Ca~{\sc ii} lines; it tends to
underestimate the EWs, thereby reducing the apparent fraction of SGs in
the coolest (M) class.

\section{Stellar bar}

The conclusion arised from this analysis is that
the observed region contains intense star formation of recent origin, since
supergiants and bright giants are necesarily young stars. 
There is a high proportion of young
stars, and their high spatial concentration shows
that the star formation has taken place in the zone observed.
Any associated early B-type or O-type stars, typical
of this kind of cluster, are not observed in this case since
their emission in $K$ is too low to reach $m_K<5$ mag for objects
far from the solar neighborhood, and the emission in the visible
is very strongly affected by dust extinction.

This star formation region does not belong to the disc nor the
spiral arms (G97); nor it is likely belonging to
the ring, as exposed in G97 and Hammersley {\it et al.} (1994) (hereafter H94).
A further convincing reason for excluding a ring, even an elliptical one,
besides the ones in those papers,
is that it would be prominent in other TMGS regions
closer to the center than $\ell$=27\arcdeg, and that
the luminosity function of the stellar excess
would be invariant, neither of which would be
 in agreement with the observations.
The TMGS star counts after subtraction of a model
disc (Wainscoat {\it et al.} 1992) and bulge (L\'opez--Corredoira
{\it et al.} 1997) are zero in off-plane directions, showing that those
disk and bulge models are good fits to the observations;
there is, however, an excess of stars in the plane ($|b|<2$\arcdeg). 
Figure \ref{Fig:residuos_l} shows a sharp decrease in the  star counts
in the center, and also  that the luminosity function of the
stellar excess at $\ell$=27\arcdeg \ is different from that in the other regions
(the ratio between stars with $m_K\le 9.0$ mag 
and $m_K \le 5.0$ mag is very different). 
These considerations lead to the conclusion that a ring-type structure,
even elongated and eccentric, gives a poor fit to the observations.

\begin{figure}[htb]
\label{Fig:residuos_l} 
\end{figure}

A hole in the extinction was suggested as a possible cause of the excess
at $\ell$=27\arcdeg \ by Jones {\it et al.} (1981) but
this idea was criticized by Kawara {\it et al.} (1982) who observed an invariant
(H-K) colour for the stars across this region. 
Also, the TMGS star counts at $\ell$=27\arcdeg \ in the plane
show an excess of over 100 stars per square degree up to $m_K=5$
and this cannot be satisfactorily explained by assuming that the extinction is zero in that region,
according to our calculations. Furthermore, CCD $VRI$ photometry 
of the region shows clearly that the line-of-sight extinction is greater 
than the standard value of 0.62 mag kpc$^{-1}$ in V (Mahoney 1999, Hammersley
{\it et al.} 1998).

The most likely explanation is the presence of a giant
star formation region which is
associated with the receding tip of the stellar Galactic bar 
(see Fig. \ref{Fig:modelores}).
If we take the region to be in the center of the star formation region
at the end of a bar\footnote{Its center is at $\ell\sim 27\arcdeg  $, $b\sim 0\arcdeg $. More 
precisely, Viallefond {\it et al.} (1980) places the region with center at $\ell=27.5\arcdeg $.
The region is extended over several degrees but this is the point
where the star counts show a maximum which we believe correspond to the center.}, 
it can be used not only as a qualitative demonstration of the bar's
existence but also as a means to estimate its orientation.
The remaining stars in Fig. \ref{Fig:residuos_l} can be explained
as bar stars, with a prominent peak at both ends due to the
interaction with spiral arms. 
Such regions can form due to the concentrations of
shocked gas where the Galactic stellar bar meets a spiral arm,
 as is observed at
the ends of the bars of face-on external galaxies (Sandage \& Bedke 1994).
An excess of extincion at negative latitudes (Calbet {\it el al.} 1996{\it b})
is also explained by the stellar-bar hypothesis.

\begin{figure}[htb]
\label{Fig:modelores} 
\end{figure}

This is not the first time that a Galactic bar has been claimed to
be discovered but our arguments in favour of it are new and different from
those of other authors.
De Vaucouleurs (1964, 1970) first suggested that the Galaxy might be barred 
in an attempt to explain the observed non-circular gas
orbits. Since then, many types of observational evidence
have been accumulated that support this hypothesis
(see Blitz 1993; Blitz et al. 1993; Kuijken 1996a; Gerhard {\it et al.} 1997). The
axial asymmetry of the inner Galaxy is detected by  
star counts (Nakada {\it et al.} 1991; Weinberg 1992; 
Whitelock {\it et al.} 1992; Stanek {\it et al.} 1994, 1996; 
W\'ozniak \& Stanek 1996; Nikolaev \& Weinberg 1997), or by surface photometry at different wavelengths
(Blitz \& Spergel 1991; Weiland {\it et al.} 1994; Dwek {\it et al.} 1995,
Sevenster 1996), stellar population studies in Baade's window 
(Ng {\it et al.} 1996), micro-lensing (Paczy\'nski {\it et al.}
1994) and studies of the extinction using lensed stars (Stanek 1995), 
analysis of internal motions of the gas
(Peters 1975; Liszt \& Burton 1980; Yuan 1984; Gerhard 1996). 
Models including a bar ({\it e.g.}, Binney {\it et al.} 1991; Zhao 1996) 
can explain these observational features. 

Not all of these observations 
necessarily imply the existence of a bar; some could be explained
by a triaxial structure in the inner Galaxy. 
Whether this structure is a triaxial bulge,  
or whether a bar or both features are present is still a matter of controversy 
(Kuijken 1996{\it b}; Ng 1997), and
there are reasons for believing that many authors who refer to a bar are 
in fact referring to the triaxial bulge.

In our previous analysis of the Two Micron Galactic Survey (TMGS) database
(Garz\'on {\it et al.} 1993) we described evidence in favor of the 
presence of a triaxial bulge (L\'opez-Corredoira {\it et al.} 1997) 
with radial scale length $\sim $2.2 kpc,
making an angle $12\arcdeg \  $ with respect to the Sun--Galactic center 
line in the first quadrant, in strong agreement with the angle 
given by many of other authors 
(Binney {\it et al.} 1991;
Weinberg 1992; Dwek {\it et al.} 1995; Binney {\it et al.} 1997; 
Stanek {\it et al.} 1997;
Nikolaev \& Weinberg 1997) whose bar angle 
(around 20 degrees) is quite close to this.
The bar of this paper must be a different structure from
the triaxial bulge since the inclination of the bulge
is not sufficient for a bar tip to reach $\ell=27\arcdeg \  $ nor for a
dust lane to reach $\ell=-19\arcdeg \  $ at the other side
of the bar, as found by Calbet {\it et al.} (1996{\it b}). 
The triaxial bulge reported by
L\'opez-Corredoira {\it et al.} (1997) does not yield sufficient
star counts in the plane at positive Galactic latitudes,
taking into account the disk and the extinction, to explain the
observations.

There is another infrared peak at negative
Galactic longitudes: at $\ell=-22\arcdeg $, which is very similar 
to the one analyzed here\footnote{See {\it COBE}-DIRBE data (Boggess {\it 
et al.} 1992)
or near-infrared catalogs which cover
the Galactic plane at positive and negative Galactic longitudes. See also
H94, and Calbet {\it et al.} (1996{\it b}).}. 
If we assume that the other end of the bar is at $\ell=-22\arcdeg $,
the orientation of the bar can be derived.
In fact, it is a simple trigonometrical problem when we take the bar
to be rectilinear and with equal length from each end to the center.

Applying the sine rule (see Fig. \ref{Fig:barangle}),

\begin{equation}
\frac{L_0}{\sin 27\arcdeg }=\frac{R_0}{\sin (180\arcdeg -27\arcdeg -\alpha)}
,\end{equation}

\begin{equation}
\frac{L_0}{\sin 22\arcdeg }=\frac{R_0}{\sin (\alpha-22\arcdeg )}
.\end{equation}

Hence, with $R_0=7.86$ kpc (L\'opez-Corredoira {\it et al.} 1997), we deduce
that $\alpha =75.6\arcdeg \  $ and the length of the bar from tip to tip
is $2L_0=7.4$ kpc.

\begin{figure}[htb]
\label{Fig:barangle}
\end{figure}

Therefore, we suspect that the authors which are giving a much
lower angle are in fact analysing the angle
of the triaxial bulge, or an average angle
of the composition of both bulge and bar, instead the bar angle.
The triaxial bulge cannot be causing the features seen at $\ell=27$\arcdeg .

We can also derive geometrically the distance to the tip
at $\ell=27\arcdeg $ as being $7.8$ kpc. This distance is compatible
with that derived in G97 (in table \ref{Tab:espectros} are the distances
for each star), taking into account all the
uncertainties in the calculation.

\section{Conclusions}

This paper, together with its first part (G97) makes a
spectroscopic analysis of the brightest stars in an infrared selected sample
of objects close to the Galactic plane at $\ell=27$\arcdeg\ showing a strikingly
high fraction of supergiants, characteristic of a strong star formation region.

We argue, using all the available evidence, that this region is situated at 
one end of the Galactic bar, where the Scutum spiral arm breaks away, and that
the presence of the region is itself a
powerful evidence for the presence of the bar.
None of the alternative possibilities (arm, disk, bulge or ring components)
is capable of explaining the observations in a satisfactory manner.

The detected presence of a similar concentration of near-IR sources in the
plane at $\ell=-22$\arcdeg\ should, on this hypothesis, indicate the other
end of the bar. To confirm this requires a similar spectroscopic campaign to that
reported here. A rectilinear bar between these two points makes
an angle of 75\arcdeg \ with the Sun--Galactic centre line, and has
a total length of 7.4 kpc. The distance this implies
to the star-forming zone analyzed here is consistent with
the estimated spectroscopic parallaxes of the stars
whose spectra we have analyzed in the first part of the paper (G97), 
giving a self--consistent picture of the bar.

It is important to distinguish this bar from the triaxial bulge
of the Galaxy (L\'opez-Corredoira {\it et al.} 1997). Many of the previous
claims to describe a Galactic bar are more likely to refer 
to the triaxial bulge and not to the bar.

\acknowledgements

The Isaac Newton Telescope is operated on the island of La Palma by the
Royal Greenwich Observatory in the Spanish Observatorio del Roque de
Los Muchachos of the Instituto de Astrof\'{\i}sica de Canarias.
This work was partially supported by grants PB94-1107, and
PB97-0219 of the
Spanish DGICYT.


\begin{figure}[htb]

\label{Fig:spectraplot} 
\end{figure}








\newpage

{\bf \Large FIGURE CAPTIONS}

\begin{description}

\item[Figure 1:] Equatorial coordinates (J2000) of the stars whose spectra 
are analyzed in this paper. Small circles stand for giants 
and big circles stand for supergiants.

\item[Figure 2:] Comparison between $EW$ measured in unmasked
Ca{\sc ii} lines using (\protect{\ref{EW}}) and the subsample having
the $EW$ obtained via (\protect{\ref{EWED}}).

\item[Figure 3:] TMGS star counts as functions of Galactic longitudes in the Galactic plane
after subtraction of the Wainscoat et al. (1992) model disc and the
L\'opez-Corredoira et al. (1997) model bulge.

\item[Figure 4:] A bar can explain a number of observations: the excess of
extinction at negative Galactic latitudes due to a dust lane (Calbet
et al. 1996b), and the star formation regions at each end of the bar.

\item[Figure 5:] Geometry of the bar sustaining an 
angle $\alpha $ with respect to the Sun (S)-Galactic center (C) line.

\item[Figure 6:] Plots of the spectra from our sample in the range
7900\AA - 8900\AA . The intrinsic slopes and those due to reddening have
been removed. Numbers for the stars in the right hand side correspond to
those given in the first column of the Table \protect{\ref{Tab:espectros}}, and the spectral
type and luminosity class of the star.

\end{description}

\begin{table*}[htb]
\begin{tabular}{cccccccc} \\ \hline
\label{Tab:espectros}
\end{tabular} 
\end{table*}

\end{document}